\documentclass[12pt]{iopart}

\usepackage{iopams}
\usepackage{graphicx}
\usepackage{caption}
\usepackage{sistyle}
\usepackage[noadjust]{cite}

\newcommand{\ba}{\begin{eqnarray}}
\newcommand{\ea}{\end{eqnarray}}

\begin{document}

\title[On the universal AC optical background in graphene]
{On the universal AC optical background in graphene}

\author{V P Gusynin$^1$, S G Sharapov$^1$ and J P Carbotte$^2$}

\address{$^1$ Bogolyubov Institute for Theoretical Physics, 14-b
        Metrologicheskaya Street, Kiev, 03680, Ukraine}
\address{$^2$ Department of Physics and Astronomy, McMaster University,
        Hamilton, Ontario, Canada, L8S 4M1}
        \ead{vgusynin@bitp.kiev.ua} \ead{carbotte@mcmaster.ca}  \ead{sharapov@bitp.kiev.ua}

\begin{abstract}
The latest experiments have confirmed the theoretically expected universal value $\pi e^2/2h$ of the ac conductivity of graphene and
have revealed departures of the quasiparticle dynamics from predictions for the Dirac fermions in idealized graphene.
We present analytical expressions for the ac conductivity in graphene which allow one to study how it is affected by
interactions, temperature, external magnetic field and the opening of a gap in the quasiparticle spectrum.
We show that the ac conductivity of graphene does not necessarily give a metrologically accurate value of the von Klitzing constant
$h/e^2$, because it is depleted by the electron-phonon interaction. In a weak magnetic field
the ac conductivity oscillates around the universal value and the Drude peak evolves into a peak at the cyclotron frequency.
\noindent
(Some figures in this article are in colour only in the electronic version)
\end{abstract}

\submitto{\NJP}
\maketitle

\section{Introduction}
\label{sec:intro}

The fascinating progress in the study of graphene
\cite{Geim2007NatMat,Neto2009RMP} in many ways is based on the usage of various experimental
methods that provide new information on the intriguing electronic properties of this material.
An impressive example is provided by the measurements of ac response which includes reflectance, transmission and
extracted from them conductivity
of monolayer graphene \cite{Wang2008Science,Nair2008Science,Li2008NatPhys}, bilayer graphene \cite{Kuzmenko2009PRB},
epitaxial graphene \cite{Dawlaty2008APL},
and graphite \cite{Kuzmenko2008PRL} reported in 2008. These measurements have made possible a quantitative  comparison of the
value of the universal ac conductivity background
extracted from the experimental data and the theoretical result
\cite{Ando2002JPSJ,Gusynin2006PRL}   predicted earlier which   at zero temperature
can be written in a particularly simple analytical form
\begin{equation}
\label{B=0.cond-intra-inter-T=0} \mbox{Re} \sigma_{xx}(\Omega)=
\frac{e^2}{h}|\mu|\frac{4 \Gamma}{\Omega^2 + 4 \Gamma^2}+
\frac{\pi
e^2}{2h}\theta\left(\Omega -2 |\mu|\right).
\end{equation}
Here $\Omega(>0)$ is the photon energy, $\mu$ is the chemical potential tunable by the gate voltage and
$\Gamma$ is the impurity scattering rate which is set to zero in the second, interband, term.
It is remarkable that the interband conductivity is universal, independent of the band structure
parameters. The dependence of the absorption edge on concentration  in semiconductors is called as the Burstein-Moss
effect. It allows one to measure directly the value of the chemical potential of the system.
For finite temperature versions of Eq.~(\ref{B=0.cond-intra-inter-T=0}) see, for example,
Refs.~\cite{Gusynin2006PRL,Falkovsky2007EPJB,Falkovsky2007PRB}.

The  theoretical value $\pi e^2/(2h)$ of the universal ac conductivity of monolayer graphene
is in good agreement with  experiments in the infrared range of wavelengths
$\lambda \simeq 1500 - \SI{2500}{nm}$ \cite{Li2008NatPhys} and even in the visible with
$\lambda \simeq 400 - \SI{700}{nm}$ \cite{Nair2008Science}. The last result seems to
be well beyond the validity of the Dirac
approximation used in \cite{Gusynin2006PRL,Falkovsky2007EPJB}, but as shown in \cite{Stauber2008aPRB}
the corrections to the Dirac-cone approximation are a few percent only.

Even earlier, similar measurements were made in an applied magnetic field on epitaxial graphene
\cite{Sadowski2006PRL,Plochocka2008PRL},  monolayer graphene  \cite{Jiang2007PRL,Deacon2007PRB},
bilayer graphene
\cite{Henriksen2008PRL} and highly oriented pyrolytic graphite \cite{Li2006PRB}.
The ac conductivity was extracted from the reflectance
in this last paper only, while in  \cite{Sadowski2006PRL,Jiang2007PRL,Henriksen2008PRL} the relative transmission
is considered. Knowledge of relative transmission allows one to investigate the magnetic field dependence of the
energy levels and find a band velocity, although as yet does not allow one to study
the redistribution of the optical spectral
weight between different energy regions as done in \cite{Kuzmenko2008PRL}.
On the theoretical side, the magneto-optical
response of graphene has been studied in Refs.~\cite{Peres2006PRB,Gusynin2006PRB,Gusynin2007PRL,Gusynin2007JPCM}, and
including the role of interactions
and collective excitations,  in Refs.~\cite{Bychkov2008PRB,Fischer2009}.

Besides the agreement with early theoretical predictions
\cite{Ando2002JPSJ,Gusynin2006PRL,Falkovsky2007EPJB,Peres2006PRB,Gusynin2006PRB,Falkovsky2007PRB},
several observations \cite{Li2008NatPhys} indicate the relevance of many-body interactions
to the electromagnetic response of graphene. Those include unusual large broadening around $2|\mu|$ which cannot be explained
by thermal smearing of the curve described by (\ref{B=0.cond-intra-inter-T=0}). Also the conductivity below the threshold of
$2 |\mu|$ does not vanish due to Pauli blocking as described by Eq.~(\ref{B=0.cond-intra-inter-T=0}),
but shows a significant residual conductivity
as strong as $0.3 \pi e^2/(2 h)$ in a wide range of energies at finite doping.

An attempt to explain these
results taking into account effects of disorder and phonons (both optical and acoustic)
was made in Refs.~\cite{Peres2008EPL,Stauber2008PRB}.
Earlier the influence of optical phonons on the ac conductivity was studied in \cite{Stauber2008JPCM}
and recently corrections due to electron-electron interaction in undoped ($\mu=0$) graphene
were discussed \cite{Mishchenko2007PRL,Herbut2008PRL,Fritz2008PRB,Sheehy2009}.

Thus one may conclude that the investigation of the ac response of graphene is already reaching the
same maturity as, for example, for high-$T_c$ superconductors (HTSC) [see Ref.~\cite{Basov2005RMP}, Secs.~IV B and C, for a review],
where the connection between conductivity and quasiparticle self-energy is exploited to extract information
about the bosonic modes and their interaction with carriers.
The case of graphene is, however, more complicated than that of HTSC, because one has to take into account
the presence of interband transitions.

In this paper we present some of our latest results on the ac conductivity
accentuating the effect of various parameters such as temperature, impurity scattering rate, doping,
magnetic field and of interactions such as the electron-phonon interaction on the value  of the universal background.

\section{Model and general expressions for conductivity}

The frequency-dependent electrical conductivity tensor
$\sigma_{\alpha \beta}(\Omega)$ is calculated using the Kubo formula
\begin{equation}\label{Kubo}
\sigma_{\alpha \beta}(\Omega)= \frac{K_{\alpha
\beta}(\Omega+i0)}{-i(\Omega+i0)}, \qquad K_{\alpha
\beta}(\Omega+i0) \equiv \frac{\langle\tau_{\alpha
\beta}\rangle}{V}+ \frac{\Pi_{\alpha \beta}^R(\Omega+i0)}{\hbar V},
\end{equation}
where the retarded  current-current correlation function is given by
\begin{equation}
\Pi_{\alpha \beta}^R(\Omega)=\int_{-\infty}^\infty dt\,e^{i\Omega
t}\Pi_{\alpha \beta}^R(t),\quad \Pi_{\alpha
\beta}^R(t)=-i\theta(t){\rm
Tr}\left(\hat{\rho}[J_\alpha(t),J_\beta(0)]\right),
\end{equation}
$V$ is the volume (area) of the system, $\hat{\rho}=\exp(-\beta
H_0)/Z$ is the density matrix  with the Hamiltonian $H_0$ in the grand canonical ensemble,
$\beta=1/T$ is the inverse temperature, $Z={\rm Tr}\exp(-\beta H_0)$
is the partition function, and $J_\alpha$ are the total paramagnetic
current operators. The diamagnetic or stress tensor $\langle
\tau_{\alpha \beta}\rangle$ in the Kubo formula (\ref{Kubo}) is a
thermal average of the diamagnetic part which is nonzero only when a tight-binding model
is considered.
Accordingly, the restricted optical conductivity sum rule
(see e.g. Refs.~\cite{Carbotte:review,Benfatto:review} for a review) in graphene has a form \cite{Gusynin2007PRB}
\begin{equation}
\label{sum-rule-sigma}
\fl
\frac{1}{\pi}\int\limits_{-\infty}^\infty
d\Omega \mathrm{ Re} \sigma_{xx}(\Omega)=
\frac{\langle\tau_{xx}\rangle}{V}  = -\frac{e^2a^2}{3\hbar^2}\int_{BZ}\frac{d^2{\bf
k}}{(2\pi)^2}\left[n_F(\epsilon(\mathbf{k})) -
n_F(-\epsilon(\mathbf{k}))\right] \epsilon(\mathbf{k}),
\end{equation}
where $\epsilon(\mathbf{k})$
is full band dispersion of graphene \cite{Wallace1947PRev}, $n_F(\omega)= [\exp((\omega-\mu)/T)+1]^{-1}$ is the Fermi
distribution function and the integration is done over the two-dimensional Brillouin zone (BZ).
At large $\Omega \sim t$, where $t$ is the nearest neighbor hopping parameter, the expression for $\sigma_{xx}(\Omega)$
has to be derived from a tight-binding model \cite{Gusynin2007PRB}. As studied in  Ref.~\cite{Stauber2008PRB}, for
such large energies  it is necessary to include in the analysis a hopping parameter for second-nearest neighbors
with a value of the order of $0.1t$.

Still, one may consider the partial sum rules in the low-energy regime \cite{Sabio2008PRB} based on the
linearized continuum approximation for the tight-binding Hamiltonian. A big advantage of this approximation
is that it allows one to obtain many transparent analytical results
that elucidate the physics of graphene. Even more such insight comes when the low-energy description of
graphene is mapped onto $2+1$-dimensional electrodynamics \cite{Semenoff1984PRL} (see Ref.~\cite{Gusynin2007IJMPB} for a review).
In many cases, the effective low-energy QED$_{2+1}$ description
provides a very good starting point for the investigation of
various transport properties making it possible to
profit from the analogies with particle physics and to
use its powerful field theoretical methods.
In this case, the low-energy quasiparticle excitations  are described
in terms of a four-component Dirac spinor $\Psi_{s}^T= \left(
\psi_{K_{+}As},\psi_{K_{+}Bs},\psi_{K_{-} Bs}, \psi_{K_{-} As}\right)$.
The spinor (with a given spin index $s=\pm$) combines the Bloch states
on the two different sublattices ($A$ and $B$) of the hexagonal graphene
lattice and with momenta near the two inequivalent Dirac points ($K_+$
and $K_-$) of the two-dimensional Brillouin zone.
The low-energy Lagrangian density for quasiparticles can be written in a relativistic-like form as
 \begin{equation}
\label{Lagrangian-M}
\fl
\mathcal{L} = \sum_{\sigma = \pm} \bar{\Psi}_\sigma (t, \mathbf{r})
\left[ i \gamma^0
\left(\hbar \partial_t - i  \mu_\sigma  \right) + i \hbar v_F  \gamma^1
D_x + i \hbar v_F  \gamma^2 D_y - \Delta \right] \Psi_\sigma (t, \mathbf{r}),
\end{equation}
where $\bar{\Psi}_{\sigma} = \Psi_{\sigma}^\dagger \gamma^0$ is the Dirac
conjugated spinor, the Fermi velocity $v_F = 3 t a_{CC}/(2 \hbar )\approx c/300$ plays the role of the speed of light $c$.
The orbital effect of a perpendicular magnetic field
$\mathbf{B} = \nabla \times \mathbf{A}$ is included via the covariant
derivative $D_\alpha=\partial_\alpha+(ie/\hbar c)A_\alpha$, where $\alpha=x,y$ and the Zeeman energy is included
via splitting
$\mu_\sigma = \mu - \sigma \mu_B B$  of the chemical potential $\mu$ with
$\mu_B = e \hbar/(2 m c)$ being the Bohr magneton.
The $4\times4$ matrices $\gamma^\nu$ furnish a reducible representation
of the Dirac algebra. Here, we use the following representation:
\begin{equation}
\label{chiral-gamma}
\gamma^0={\tilde \tau}_1 \otimes \tau_0,\qquad
\gamma^\alpha=-i {\tilde \tau}_2 \otimes \tau_\alpha,
\end{equation}
where the Pauli matrices $\tilde{\tau}_\alpha$ and $\tau_\alpha$ (as well as the $2\times2$
unit matrices $\tilde{\tau}_0$ and $\tau_0$) act on the valley ($K_+,\, K_-$) and
the sublattice ($A,\, B$) indices, respectively. This representation is derived
from the tight-binding model for graphene and to make the treatment more general we also include a
mass (gap) term with $\Delta$ (see e.g. review \cite{Gusynin2007IJMPB}).
Accordingly, the Kubo formula (\ref{Kubo}) can now be calculated
using the electric current density operator  $j_\alpha(t,\mathbf{r})$ that follows from the Lagrangian (\ref{Lagrangian-M}):
\begin{equation}
\label{current-Dirac}
\fl
J_{\alpha}(t)=\int d^{2}r j_{\alpha}(t,\mathbf{r} ), \qquad
j_\alpha(t,\mathbf{r}) = \frac{\delta \mathcal{L}}{\delta (A_\alpha/c)}=
-ev_{F}\sum_\sigma\bar{\Psi}_\sigma(t,\mathbf{r})\gamma^\alpha \Psi_\sigma(t,\mathbf{r}).
\end{equation}
All results in this paper are obtained
neglecting the vertex corrections. In this case the calculation of the conductivity reduces to the
evaluation of the bubble diagram. The corresponding polarization operator can
be written in the form
\begin{eqnarray}
\label{Pi-via-RA}
& & \frac{\Pi_{\alpha \beta}  (\Omega+i0)}{V}
=-\frac{e^2v_F^2}{2\pi i}\int\limits_{-\infty}^\infty
d\omega\,n_F(\omega)\int\frac{d^2k}{(2\pi)^2}  \\
& &\times {\rm tr} \left[\gamma^\alpha (G^R  (\omega,\mathbf{k})
-G^A   (\omega,\mathbf{k}))\gamma^\beta (G^A  (\omega-\Omega,\mathbf{k})
+G^R (\omega+\Omega,\mathbf{k}))\right] \nonumber,
\end{eqnarray}
where the $\mbox{tr}$ also includes the summation over spin index and
$G^{R(A)}  (\omega,\mathbf{k})$ are retarded (advanced) fermion Green's functions (GF).
The approximation of Eq.~(\ref{Pi-via-RA}) does not correctly reproduce the
Boltzmann dc conductivity because of the omission of
vertex corrections. The current vertex corrections for
electron-phonon scattering in graphene within the Kubo formalism for the dc conductivity were considered
in Ref.~\cite{Cappelluti2009PRB}. For the Coulomb interaction and ac conductivity the vertex corrections were considered
in Ref.~\cite{Sheehy2009}.
In the absence of magnetic
field these GF are given by (we also set  $\hbar =1$, so that $\Omega$ can be regarded both as the frequency and energy of the photon)
\begin{equation}
\label{GF}
G^{R, A}  (\omega,\mathbf{k}) = \frac{\gamma^0 (\omega - \Sigma^{R,A} (\omega) )- v_F \mathbf{k} \pmb{\gamma} + \Delta}
{(\omega - \Sigma^{R,A} (\omega) )^2 - v_F^2 \mathbf{k}^2 -\Delta^2}.
\end{equation}
In (\ref{GF}) we included the retarded (advanced) electronic self-energy $\Sigma^{R (A)} (\omega)$
which represents the contribution due to impurities, phonons or electron-electron interactions and
which in general also depends on the wave vector $\mathbf{k}$. However, because we
study self-energy corrections due to optical phonons in what follows we omit this dependence.
We write all expressions in terms of the retarded self-energy $\Sigma^{R} (\omega)$ related to the advanced by a
complex conjugation $\Sigma^{A} (\omega) = [\Sigma^{R} (\omega)]^*$,
so that the subscript ``R'' will be dropped.

The calculation of the conductivity follows closely that described in Appendix~B of \cite{Gusynin2006PRB}, so we
present a final result for the case of zero magnetic field
\begin{eqnarray} \label{sigma_ac}
\fl  \sigma_{xx} && (\Omega) =\frac{e^2}{\pi^2\Omega}{\rm
Re}\int\limits_{-\infty}^\infty
d\omega[n_F(\omega)-n_F(\omega+\Omega)] \nonumber \\
\fl  && \times \left[\left( \frac{(\omega - \Sigma(\omega))(\omega^\prime - \Sigma(\omega^\prime))-
\Delta^2}{[\Sigma(\omega^\prime) -\Sigma(\omega)-\Omega]
[2 \omega + \Omega - \Sigma(\omega) - \Sigma(\omega^\prime)]} -\right.\right.\\
\fl && \left.\left.\frac{(\omega -\Sigma(\omega) )(\omega^\prime- \Sigma^\ast(\omega^\prime))-\Delta^2}
{[\Sigma^\ast(\omega^\prime) - \Sigma(\omega) - \Omega]
[2\omega + \Omega - \Sigma(\omega) - \Sigma^\ast(\omega^\prime)]}\right)\ln[{\Delta^2-(\omega - \Sigma(\omega))^2}]+ (\omega\leftrightarrow\omega^\prime)\right], \nonumber
\end{eqnarray}
where $\omega^\prime = \omega + \Omega$. The expression (\ref{sigma_ac}) can be explicitly calculated
as was done in Ref.~\cite{Gusynin2006PRB} for $\mbox{Re} \Sigma(\omega) =0$  and in Ref.~\cite{Stauber2008PRB} for $\Delta=0$
including both $\mbox{Re} \Sigma(\omega)$ and $\mbox{Im} \Sigma(\omega)$. These final expressions are somewhat cumbersome as
compared to  the representation given in (\ref{sigma_ac}).

\section{AC conductivity in the absence of an external magnetic field and electron-phonon interaction}
\label{sec:conductivity-B=0}

As mentioned in Sec.~\ref{sec:phonons}, for the correct analysis of the ac conductivity with
phonons it is crucial to include both
real and imaginary part of the self-energy $\Sigma(\omega)$. However, to gain some first insight
into Eq.~(\ref{sigma_ac})
it is useful to consider first the case $\mbox{Re} \Sigma(\omega)=0$. Considering the  imaginary part of the self-energy,
$\Gamma(\omega) = - \mbox{Im} \Sigma(\omega)$ to be small, one can obtain the following representation
\cite{Gusynin2006PRL} for the conductivity
\begin{eqnarray}
\label{B=0.cond-intra-inter}
\fl \mbox{Re} \sigma_{xx}(\Omega,T) &=& \frac{e^2}{\pi^2
\hbar}\int\limits_{-\infty}^\infty
d\omega\frac{[n_F(\omega)-n_F(\omega^\prime)]}{\Omega}\frac{\pi}{4\omega\omega^\prime}  \\
&\times& \left[\frac{2\Gamma(\omega)}{\Omega^2+4\Gamma^2(\omega)}-\frac{2\Gamma(\omega)}
{(\omega+\omega^\prime)^2+4\Gamma^2(\omega)}\right]
(|\omega|+|\omega^\prime|)(\omega^2+{\omega^\prime}^2), \nonumber
\end{eqnarray}
where  the first term in square brackets of
Eq.~(\ref{B=0.cond-intra-inter}) describes the intraband transitions
and the second term describes the interband transitions. In the clean
limit, $\Gamma\to0$, both terms in the square brackets reduce to delta
functions and we obtain a simple analytical expression,
\begin{eqnarray}
\fl \mbox{Re}\, \sigma_{xx}(\Omega) = \frac{4 e^{2}}{h}\frac{ 2 T \Gamma}{\Omega^2 + 4 \Gamma^2}
\ln \left(2 \cosh \frac{\mu}{2T} \right)
+\frac{\pi e^2}{2 h}\left[n_{F}\left(-\frac{\Omega}{2}\right)-n_{F}\left(\frac{\Omega}{2}\right)\right],
\label{sigmaxx:cleanlimit}
\end{eqnarray}
where as in Eq.~(\ref{B=0.cond-intra-inter-T=0}) the impurity scattering rate is only kept
in the intraband (Drude) term
for a regularization. In the limit $\Gamma \to 0$ the real part of the intraband conductivity can be written as a Dirac
delta-function
\begin{equation}
\label{sigma-inter}
\mbox{Re}\,\sigma_{xx}^{intra}(\Omega) = \frac{4\pi e^2 T}{h} \ln \left[2 \cosh
\left( \frac{\mu}{2T} \right) \right] \delta(\Omega).
\end{equation}
The analysis of experimental data \cite{Dawlaty2008APL,Li2008NatPhys} created a need
for various generalizations of Eq.~(\ref{B=0.cond-intra-inter-T=0}) to finite temperature,
finite impurity scattering rate in the interband term and  a nonzero gap $\Delta$.
Postponing the analysis of formulae with nonzero gap to the end of the section, we rewrite
Eq.(\ref{sigmaxx:cleanlimit}) in a slightly different form
 (see also Refs.~\cite{Falkovsky2007EPJB,Falkovsky2007PRB} and a review \cite{Falkovsky2008UFN}),
\begin{eqnarray}
\label{B=0.cond-intra-inter-T}
\sigma_{xx}(\Omega)= \sigma_{xx}^{intra}(\Omega) + \sigma_{xx}^{inter}(\Omega)\nonumber  \\
\mbox{Re}\, \sigma_{xx}^{intra}(\Omega) = \frac{e^2}{h}\frac{ 8 T \Gamma}{\Omega^2 + 4 \Gamma^2}
\ln \left(2 \cosh \frac{\mu}{2T} \right), \\
\mbox{Re}\, \sigma_{xx}^{inter}(\Omega)= \frac{\pi e^2}{2 h} \frac{\sinh (\Omega/2T)}{\cosh (\mu /T)
+ \cosh(\Omega/2T) }, \nonumber
\end{eqnarray}
For a finite $T$ and small $\Omega$ the interband term in Eq.~(\ref{B=0.cond-intra-inter-T})
is linear in $\Omega$ with decreasing slope as $\mu$ is increased. For $\Omega \gg |\mu|$ and
$\Omega/T \gg 1$ we recover the universal background value.
So finite temperature effects do not deplete this background.
\begin{figure}[h]
\centering{
\includegraphics[width=0.48\textwidth]{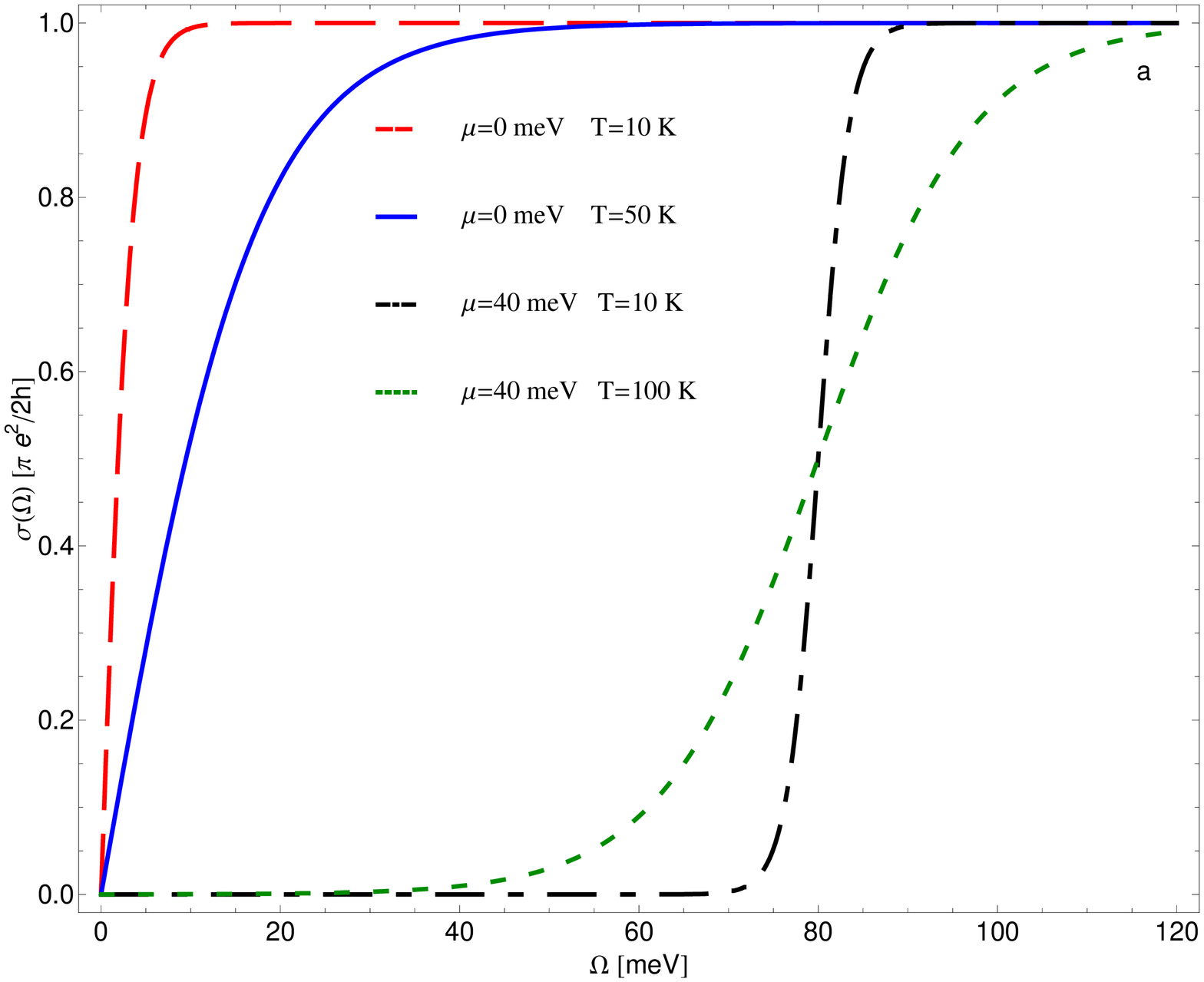}
\includegraphics[width=0.48\textwidth]{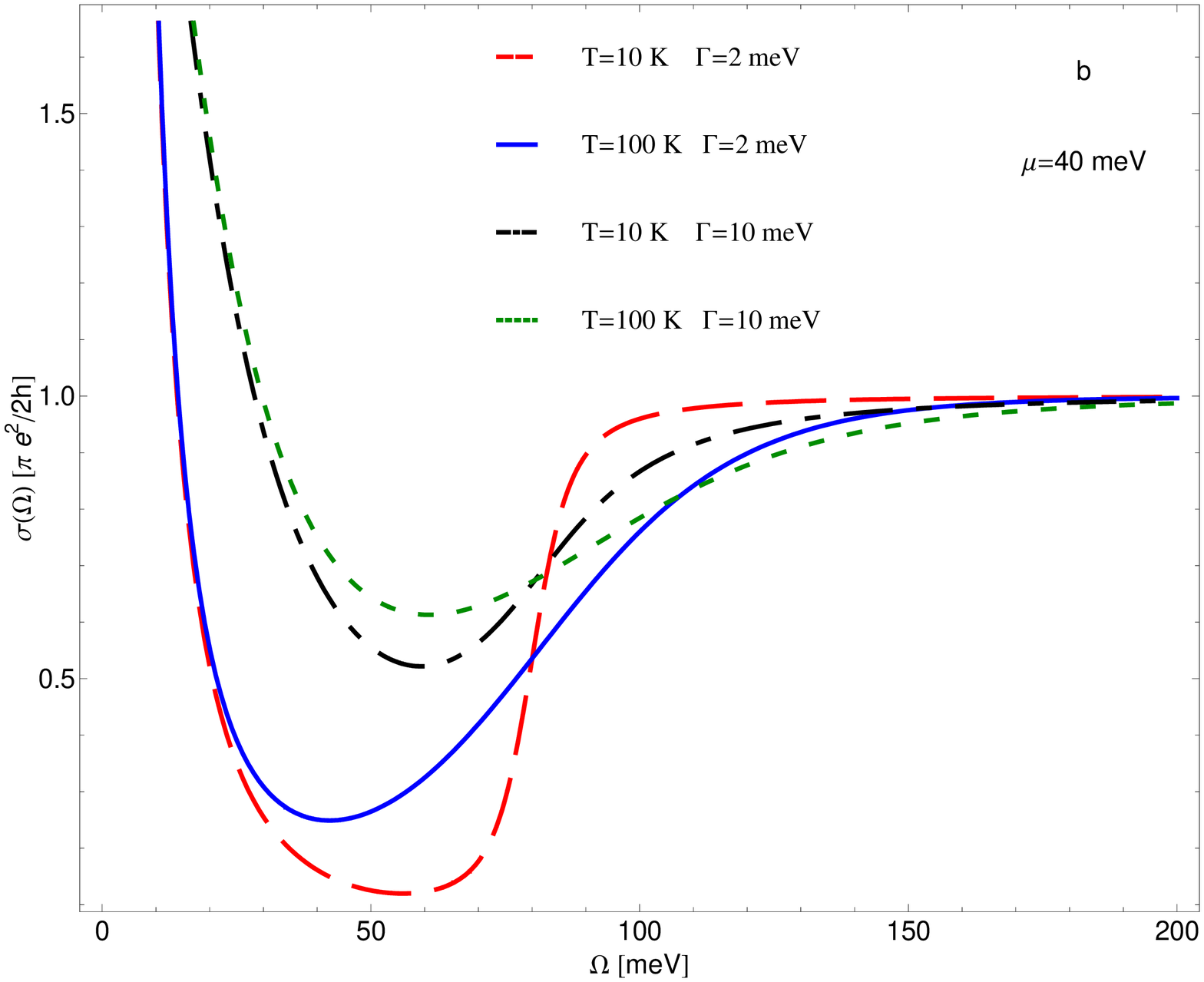}}
\caption{(a) The interband  conductivity $\mbox{Re} \sigma_{xx}^{inter}(\Omega)$ in units $\pi e^2/2 h$ (universal background):
for $\mu = \SI{0}{meV}$ the long dashed (red) line at $T = \SI{10}{K}$, the solid (blue) line
at $T = \SI{50}{K}$, and for $\mu = \SI{40}{meV}$  dash-dotted (black) line at $T = \SI{10}{K}$,  short dashed
(green) line at $T = \SI{50}{K}$. There are no  interactions and the impurity scattering is set to zero.
(b) The  conductivity $\mbox{Re} \sigma_{xx}(\Omega)$ in units $\pi e^2/2 h$
for $\mu = \SI{40}{meV}$. The long dashed (red) line at $T = \SI{10}{K}$, the solid (blue) line
at $T = \SI{50}{K}$ both are for $\Gamma = \SI{2}{meV}$. The  dash-dotted (black) line at $T = \SI{10}{K}$,  short dashed
(green) line at $T = \SI{50}{K}$ are for $\Gamma = \SI{10}{meV}$. }
\label{fig:1}
\end{figure}
These features are shown in Fig.~\ref{fig:1}a where we show numerical results for the interband conductivity
at two temperatures $T = \SI{10}{K}$ and $T = \SI{50}{K}$ for the neutrality point, $\mu = 0$,
and for $\mu = \SI{40}{meV}$. We notice the depletion below $\pi e^2/2h$ of the low frequency
region with increasing temperature and
particularly with increasing value of the chemical potential. At large frequency $\Omega$
the background remains unaltered.
In all cases, however, the missing spectral weight is transferred to the intraband Drude
term not shown in the figure. To establish this we show that a sum rule applies to the spectral
weight under the real part of
the conductivity.
Integrating  $\mbox{Re} \sigma_{xx}(\Omega)$ up to $\Lambda$ chosen to be $\Lambda \gg |\mu| \gg T$ we find
\begin{equation}
\label{partial-sum}
\int_0^{\Lambda} \mbox{Re} \sigma_{xx}(\Omega) d \Omega = W_{intra} + W_{inter}=  \frac{\pi e^2}{2 h} \Lambda,
\end{equation}
where the intraband and interband spectral weights are
\begin{equation}
\label{spectral-weights}
\fl
W_{intra}=\frac{2\pi e^{2}}{h}T\ln\left(2\cosh\frac{\mu}{2T}\right),\qquad
W_{inter}= \frac{\pi e^2}{h} T \ln\left(\frac{\cosh \Lambda/2T + \cosh (\mu/T)}{1+ \cosh (\mu/T)} \right).
\end{equation}
For $\Lambda \gg |\mu| \gg T$  we arrive at the last equality in Eq.~(\ref{partial-sum}).
This value is exactly equal to its zero temperature and zero $\mu$ value, when  $\mbox{Re} \sigma_{xx}(\Omega)$ is simply
equal to $\pi e^2/2 h$ for all $\Omega$. Our results show that the interband contribution to the dc conductivity
is zero for any finite temperature. However, there is also an intraband Drude contribution which in the clean limit
gives an infinite contribution to the conductivity [see Eq.~(\ref{sigma-inter})]. This agrees with the physical expectation  that $\sigma_{dc}$
should be infinite at any finite $T$ even at the neutrality point $\mu=0$.
There are always thermally excited electrons and holes
present at finite $T$ and these do not scatter in the clean limit, so the conductivity will be infinite and not zero as
claimed by  Peres and Stauber \cite{Peres2008IJMPB}.

In Fig.~\ref{fig:1}b we show results for the case of increasing impurity scattering $\Gamma$. In both cases shown the chemical
potential is taken to be $\SI{40}{meV}$ and two values of the temperature are used $T = \SI{10}{K}$ and $T = \SI{100}{K}$.
The long-dashed (red) curve and solid (blue) curve are for $\Gamma = \SI{2}{meV}$,
while the dash-dotted (black) curve and  short dashed (green) are for $\Gamma = \SI{10}{meV}$. What is important to note is that for
$\Omega \geq \SI{200}{meV}$ all curves have risen to the universal background value, equal to one in our units.
The width of the absorption edge at $\Omega = 2 \mu = \SI{80}{meV}$ depends both on $T$ and $\Gamma$. For example,
for the same temperature larger $\Gamma$ corresponds to a wider edge.

\subsection{AC conductivity in the presence of a gap}
\label{sec:gap}

Next we consider the possibility of the opening of a gap $\Delta$.
The most general expressions for the complex conductivity which include a finite $\Delta$
were derived in Ref.~\cite{Gusynin2007JPCM} [Eqs. (13) and (14)] then rederived in  Ref.~\cite{Dawlaty2008APL} and used to
analyze conductivity data in epitaxial graphene (see also Ref. \cite{Benfatto2008PRB} for an alternative scenario which
preserves the linear dispersion law of carriers in graphene).
In this case the clean limit generalization of
Eq.~(\ref{B=0.cond-intra-inter-T}) is
\begin{eqnarray}
\label{B=0.cond-intra-inter-T-gap}
\fl  \mbox{Re} \sigma_{xx}^{intra}(\Omega) = \frac{2 \pi e^2}{h} \delta(\Omega)
\int_{-\infty}^\infty d\omega(-n_F^\prime(\omega))\frac{(\omega^2-\Delta^2) \theta(\omega^2-\Delta^2)}{|\omega|}, \\
\fl \mbox{Re} \sigma_{xx}^{inter}(\Omega)= \frac{\pi e^2}{2 h} \frac{\sinh (\Omega/2T)}{\cosh (\mu /T) + \cosh(\Omega/2T) }
\frac{\Omega^2+4\Delta^2}{\Omega^2}\theta(\Omega^2-4\Delta^2). \nonumber
\end{eqnarray}
In the $T=0$ limit the last expression reduces to a result given in Ref.~\cite{Gusynin2006PRL}
\begin{equation} \label{B=0.cond-Delta-T=0}
\fl \sigma_{xx}(\Omega)=\frac{2\pi
e^2}{h}\delta(\Omega)\frac{(\mu^2-\Delta^2)\theta(\mu^2 -
\Delta^2)}{|\mu|} +\frac{\pi
e^2}{2h}\frac{\Omega^2+4\Delta^2}{\Omega^2}\theta\left(\Omega
-2 {\rm max}(|\mu|,\Delta)\right).
\end{equation}
The first term in Eq.~(\ref{B=0.cond-Delta-T=0}) gives the modified intraband Drude contribution and the second is the
modified interband background. It is clear that for $\Omega \gg |\mu|,\Delta$
this background again reduces to its universal value.

\section{Optical conductivity in the presence of electron-phonon interaction}
\label{sec:phonons}
\begin{figure}[h]
\centering{
\includegraphics[width=0.48\textwidth]{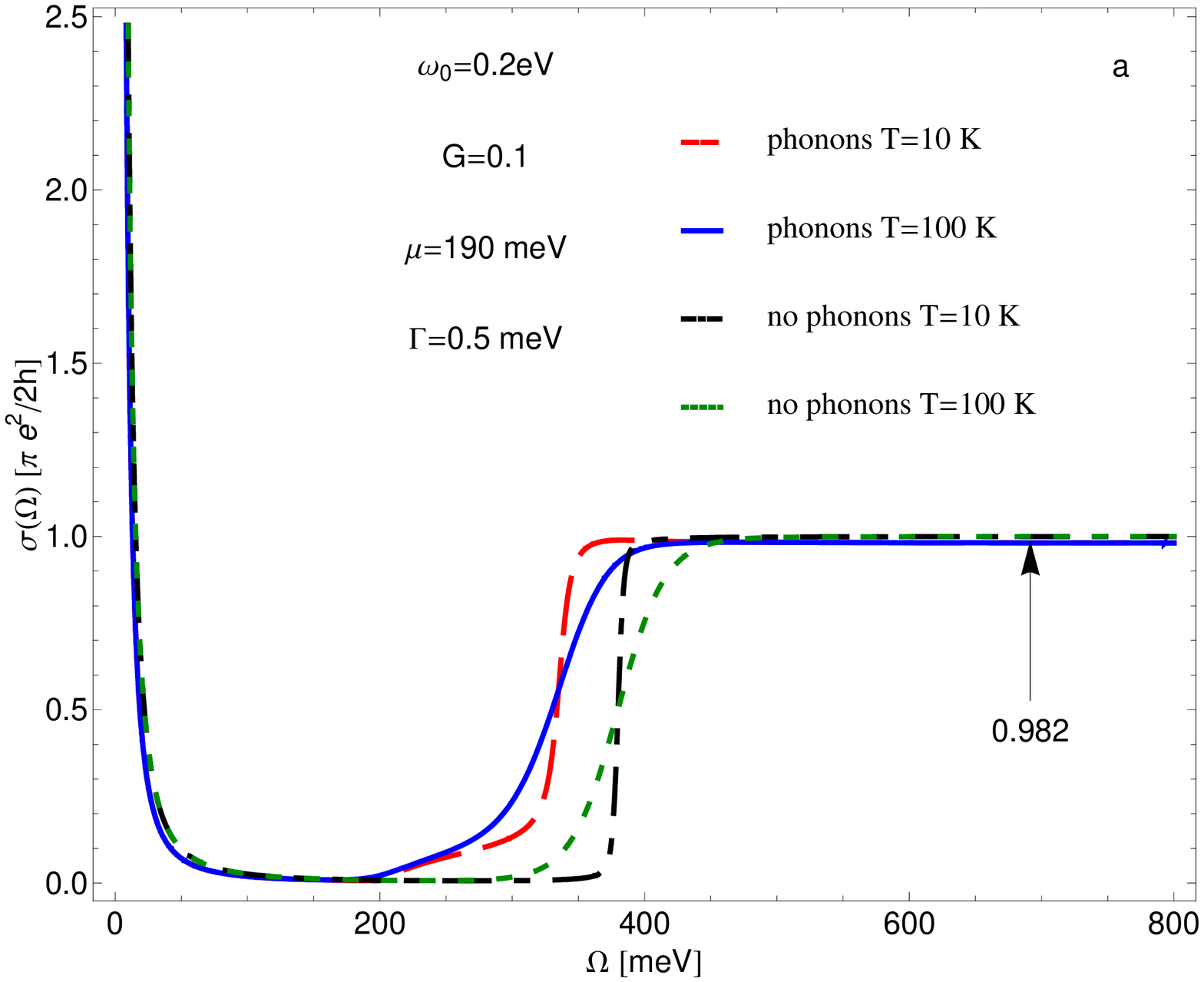}
\includegraphics[width=0.48\textwidth]{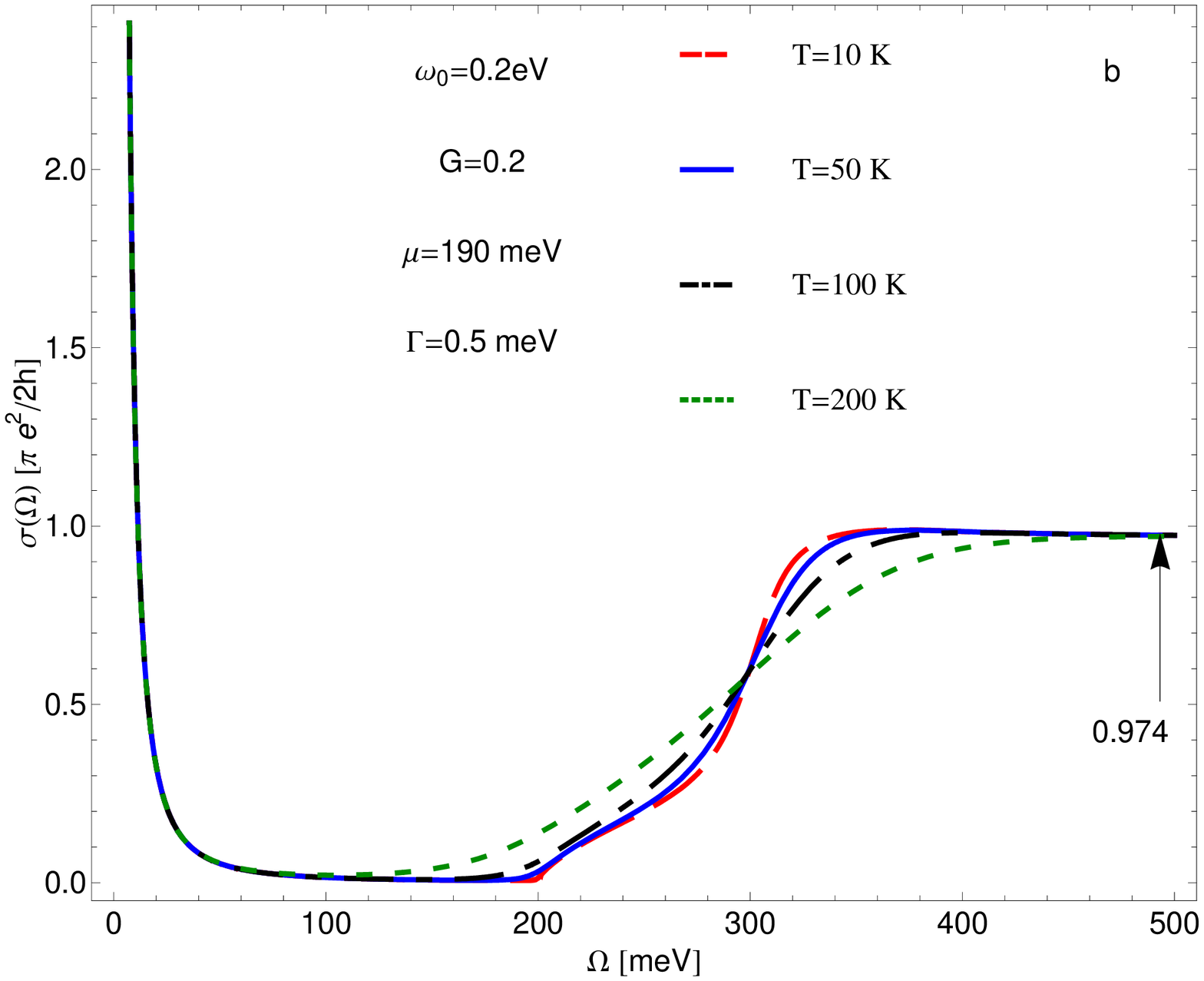}}
\caption{(a) The conductivity $\mbox{Re} \sigma_{xx}(\Omega)$ in units $\pi e^2/2 h$ (universal background)
for bare band:  dash-dotted (black) curve for $T= \SI{10}{K}$ and short dashed (green) line for $T= \SI{100}{K}$.
The curves with phonons ($\omega_0 = \SI{0.2}{eV}$ and $G=0.1$) are the long dashed (red) line for
$T= \SI{10}{K}$ and the solid (blue) line
for $T= \SI{100}{K}$. In all cases the chemical potential $\mu = \SI{190}{meV}$ and the scattering rate $\Gamma = \SI{0.5}{meV}$.
(b)
The conductivity with phonons same as (a), except to $G=0.2$, at various temperatures:
$T = \SI{10}{K}$ -- the long dashed (red) line,  $T = \SI{50}{K}$ -- the solid (blue) line,
$T = \SI{100}{K}$ -- the dash-dotted (black) line and $T = \SI{200}{K}$ -- the short dashed (green) line.
}
\label{fig:2}
\end{figure}

One may notice that the expression for the GF (\ref{GF}) does not contain the
chemical potential. This is because during the  derivation of Eq.~(\ref{Pi-via-RA})
the replacement $\omega \to \omega - \mu$
moves $\mu$ from the GF to the Fermi function $n_F(\omega)$, so that in the GF (\ref{GF})
the value $\omega=\mu$ corresponds to the
Fermi level. In the usual many-body calculations $\mu$ is kept in the GF. This implies that when
the corresponding self-energy  is inserted in the GF (\ref{GF}),
its dependence on energy has to be accordingly shifted  $\omega \to \omega - \mu$.
For example, for the simplified single-parameter model proposed in \cite{Park2007PRL}
that captures the qualitative features of
the e-ph interaction in graphene, the imaginary part of the self-energy
reads
\begin{equation}\label{our-ImSigma-phonon}
\mbox{Im} \Sigma (\omega,\mu) = - G |\omega - \mbox{sgn} (\omega - \mu)
\omega_{0}| \, \theta[(\omega -\mu)^2 - \omega_{0}^2].
\end{equation}
Here $\omega_0 \approx \SI{0.2}{eV}$ is the energy of the in-plane optical phonon mode and
$G$ is the dimensionless parameter determined by matching the model self-energy with the full {\it ab initio\/} result \cite{Park2007PRL}.
The corresponding real part of the self-energy can be obtained from Eq.~(\ref{our-ImSigma-phonon}) using Kramers-Kr\"{o}nig relations
or directly calculating the self-energy for Dirac quasiparticles coupled
to a single Einstein phonon \cite{Stauber2008JPCM,Calandra2007PRB,Tse2007PRL}
\begin{equation}
\label{our-ReSigma-phonon}
\mbox{Re} \Sigma(\omega,\mu) = \left\{
\begin{array}{lc}
 \frac{G}{\pi} \left[ \omega_{0} \ln
\left| \frac{(\omega + \omega_0)^2}{(\omega - \mu)^2 - \omega_0^2}
\right| - \omega \ln \left| \frac{W^2 (\omega - \mu +
\omega_0)}{(\omega + \omega_0)^2 (\omega - \mu - \omega_0)} \right|
\right] , & \mu>0,\\
\frac{G}{\pi} \left[ \omega_0 \ln \left| \frac{(\omega -\mu)^2 -
\omega_0^2}{(\omega -\omega_0)^2}\right| -\omega \ln \left|
\frac{W^2 (\omega - \mu- \omega_0)}{(\omega-\omega_0)^2 (\omega -\mu
+ \omega_0)}\right| \right] , & \mu <0.
\end{array}
\right.
\end{equation}
Here $W = \sqrt{\pi \sqrt{3} t} \approx \SI{7}{eV}$ is the energy cutoff
which preserves the number of states, with $t$ being the nearest neighbor hopping parameter.
By definition, the carrier effective mass renormalization $\lambda_{eff}$ due the electron-phonon interaction
is given by
\begin{equation}
\label{mass-renor-def}
\mbox{Re} \Sigma(\omega,\mu) =  - \lambda_{eff} (\omega-\mu)+ \mbox{Re} \Sigma^R(\omega =\mu,\mu),
\end{equation}
where from (\ref{our-ReSigma-phonon}) we find that
\begin{equation}
\label{mass-renorm}
\lambda_{eff}= \frac{2G}{\pi} \left(\ln \frac{W}{|\mu + \omega_0|} - 1 + \frac{|\mu|}{\omega_0}\right).
\end{equation}
The role of $\lambda_{eff}$ will be clarified in what follows. Choosing  the value of $G=0.1$ as done in Fig.~\ref{fig:2}a
below, for $\mu =0$ one obtains $\lambda_{eff} \approx 0.16$. This
is of order of the theoretical estimate \cite{Park2007PRL} and smaller than the value found experimentally
\cite{Bostwick2007Nature,Li2009PRL}.

So far we did not distinguish the bare chemical potential $\mu$ of noninteracting system
from the chemical potential $\tilde \mu$ of the interacting system. The value of $\mu$ is set by the
doping $\rho = \mbox{sgn} (\mu) \mu^2/\pi \hbar^2 v_F^2$ [here $\hbar$ is restored]
which is controlled by the gate voltage. In the interacting system the
dispersion law of the quasiparticle excitations is determined by the pole of the GF (\ref{GF}). For $\Delta =0$ and neglecting
the imaginary part of the self-energy one gets
\begin{equation}
\label{dispersion1}
\omega - \mbox{Re} \Sigma(\omega,\mu) = \pm v_F |\mathbf{k}|,
\end{equation}
or counting the energy $E$ of the excitations from the true Fermi surface of the interacting system, $\omega \to E + \tilde \mu$,
we obtain from Eq.~(\ref{dispersion1}) that
\begin{equation}
E + \tilde \mu - \mbox{Re} \Sigma (E + \tilde \mu,\tilde \mu) = \pm v_F |\mathbf{k}|,
\end{equation}
where in the vicinity of Fermi surface $\mbox{Re} \Sigma(E + \tilde \mu, \tilde \mu) = - \lambda_{eff}E + \mbox{Re}
\Sigma(\tilde \mu, \tilde \mu)$ as
follows from Eq.~(\ref{mass-renor-def}). Hence, we arrive at
\begin{equation}
\label{dispersion2}
E (1 + \lambda_{eff}) + \tilde \mu -  \mbox{Re} \Sigma (\tilde \mu, \tilde \mu) = \pm v_F |\mathbf{k}|
\end{equation}
which allows us to identify the quantity
\begin{equation}
\label{mu-eq}
\tilde \mu -  \mbox{Re} \Sigma (\tilde \mu,\tilde \mu) = \mu
\end{equation}
as the bare chemical potential of  the noninteracting system. Due to the smallness of $G$  we can assume that the
self-energy is also small and solve the equation (\ref{mu-eq}) by iterations
\begin{equation}
\label{mu-renormalization}
\tilde{\mu} = \mu + \mbox{Re} \Sigma (\mu,\mu).
\end{equation}
Thus, as discussed in Ref.~\cite{Carbotte2009} (see also references therein),
in the presence of the electron-phonon interaction one should take into account the renormalization of the noninteracting chemical
potential $\mu$ to its interacting value $\tilde{\mu}$. In the  numerical results presented below we fix the value of $\mu$ and then
in accordance with Eq.~(\ref{mu-renormalization}) compute the value of $\tilde \mu$ which has to be used as a true chemical potential
of the interacting system. As one can see from Eq.~(\ref{dispersion2}),
for the massless carriers in graphene the role of $\lambda_{eff}$ is to renormalize their velocity: $v_F \to v_F/(1+ \lambda_{eff})$
and the value of chemical potential $\mu \to  \mu /(1+ \lambda_{eff}) \approx \tilde \mu$. Restoring the imaginary part of the self-energy,
$\Gamma = - \mbox{Im} \Sigma$
in (\ref{dispersion2}), we find that the quasiparticle width is also renormalized: $\Gamma \to \Gamma /(1+ \lambda_{eff})$.
These properties allow us to use previously obtained expressions with the corresponding renormalizations.

Fig.~\ref{fig:2}a compares results for the ac conductivity $\mbox{Re} \sigma_{xx}(\Omega)$ as a function of $\Omega$
to $\SI{800}{meV}$ with and without phonons at two temperatures $T= \SI{10}{K}$ and $T= \SI{100}{K}$. For $T= \SI{10}{K}$
the dash-dotted (black) curve was calculated using bare bands, a constant residual scattering rate $\Gamma = \SI{0.5}{meV}$
and a chemical potential $\mu = \SI{190}{meV}$. This curve is to be compared with the long dashed (red) curve which
includes phonons with $G=0.1$, $\omega_0 = \SI{0.2}{eV}$ and is at the same temperature. To regularize computation
when the self-energy $\mbox{Im} \Sigma(\omega,\mu)$ given by (\ref{our-ImSigma-phonon}) is zero we use the
same value of the residual scattering rate $\Gamma = \SI{0.5}{meV}$.
While the main rise in the conductivity to its universal background is at $2 \mu = \SI{380}{meV}$ in the bare band case,
the electron-phonon renormalization according to Eq.~(\ref{mu-renormalization}) have shifted this rise to $2 \tilde \mu$, the renormalized
chemical potential which falls at considerably lower energy. Note also that at higher energies the bare
conductivity has risen to its universal value, while the renormalized one falls below, i.e. is slightly less than one
in the chosen units. Increasing the coupling $G$ reduces the background value even more as can be seen in
Fig.~2 of Ref.~\cite{Stauber2008JPCM}.
We find that for $G=5$ (which is certainly larger than expected for graphene) the conductivity at $\Omega = \SI{800}{meV}$
has dropped to a value of $0.87$. This reduction is expected on physical grounds. Phonons shift optical spectral weight
to energies beyond the bare band cutoff on the scale of a few phonon energies. This spectral weight must come from lower
energy to conserve the optical sum. We find the depletion starts for $\Omega > |\mu|$ and shows no structure.
There are no wiggles (excess conductivity associated with the Stokes and anti-Stokes lines) of the kind
reported in Refs.~\cite{Peres2008EPL,Stauber2008PRB}.
Our result agrees with an earlier observation made in
Ref.~\cite{Stauber2008JPCM} that these features disappear when the real part (\ref{our-ReSigma-phonon}) of the
self-energy is included. This is further confirmed in \cite{Peres2009EPL}.
The reduction in universal conductivity found here is important, because it shows that
the ac conductivity of graphene does not necessarily give
metrologically  accurate value of the von Klitzing constant $h/e^2$, although it does give a good estimate.
A second feature of the long dashed (red) dressed curve as compared with dash-dotted (black) curve for bare bands is the region
starting at $\Omega = \omega_0$ and extending to around $2 \tilde \mu$ (twice the dressed chemical potential).
The bare case is totally flat and small in this region, but the dressed curve shows additional absorption starting with
a clear ``kink'' at $\Omega = \omega_0$ and increasing substantially as $\Omega$ grows towards $2 \tilde \mu$.
These are the Holstein boson assisted absorption processes which proceed with the creation of a hole-particle pair
and in addition a phonon is present in the final state. These processes are well known in ordinary
metals and correspond to the incoherent part of the electronic spectral density brought about through coupling to the phonons.
At the same time the coherent Drude like contribution centered at $\Omega =0$ is reduced by a factor $1+ \lambda_{eff}$ and
contains only a fraction $1/(1+ \lambda_{eff})$ of the optical spectral weight, while the remainder
$\lambda_{eff}/(1+ \lambda_{eff})$ is to be found in the incoherent phonon assisted Holstein side band \cite{Grimvall.book}.
The short dashed (green)  and solid (blue) curve give results at $T= \SI{100}{K}$. First note that temperature does not significantly
change the value of $\mbox{Re} \sigma_{xx}(\Omega)$ at large $\Omega$ and so the ac background value is robust
with respect to temperature changes. In the region of the main rise around the bare (for the short dashed curve) and renormalized
(for solid curve) chemical potential, however, temperature does provide smearing. For the bare band this smearing is more symmetric
about twice the chemical potential than it is for the phonon renormalized case, where the signature of a phonon
contribution remains as extra absorption in the region above the phonon energy $\omega_0$ and below $2 \tilde \mu$.
Comparison with the the long dashed (red) curve at $T =\SI{10}{K}$ with that at  $T =\SI{100}{K}$ shows that, the phonon structures are
themselves smeared by temperature and are best seen at low temperature.

This is seen better in Fig.~\ref{fig:2}b where we show the phonon renormalized value of $\mbox{Re} \sigma_{xx}(\Omega)$
for four temperatures, namely, $T = \SI{10}{K}, \SI{50}{K}, \SI{100}{K}, \SI{200}{K} $. By $T= \SI{200}{K}$ the
short dashed (green) curve has become rather smooth and no sharp phonon structure remains, but smearing about $\Omega =2 \tilde \mu$
is nevertheless still quite asymmetric and in that sense retains information on phonon renormalization.

There is another way to see phonon renormalization in $\mbox{Re} \sigma_{xx}(\Omega)$ curves that is worth commenting on.
It was shown in \cite{Carbotte2009} that at $T=0$ for $|\mu| > \Omega$ (photon energy) and small residual scattering rate
$\Gamma$ the low frequency conductivity is dominated by the intraband transitions and that $\mbox{Re} \sigma_{xx}(\Omega)$
takes on a Drude like profile with a single change over the bare band case, namely,
\begin{equation}
\label{Drude-phonon}
\mbox{Re} \sigma_{xx}(\Omega) = \frac{e^2}{h} |\mu| \frac{4 \Gamma}{\Omega^2(1 + \lambda_{eff})^2 + 4 \Gamma^2 }
\end{equation}
with the phonon renormalization $\lambda_{eff}$ as defined in Eq.~(\ref{mass-renor-def}) and in the lowest approximation
given by Eq.~(\ref{mass-renorm}).  Defining the ratio of renormalized to unrenormalized conductivity
as $R(\Omega,T)$ we get at zero temperature
\begin{equation}
\label{R-ratio}
R(\Omega,T=0) = \frac{\Omega^2 + 4 \Gamma^2}{\Omega^2(1 + \lambda_{eff})^2 + 4 \Gamma^2}
\end{equation}
which starts at value 1 for $\Omega=0$ and saturates to a value of $1/(1+\lambda_{eff})^2$ at $\Omega \gg 2 \Gamma$.
\begin{figure}[h]
\centering{
\includegraphics[width=0.48\textwidth]{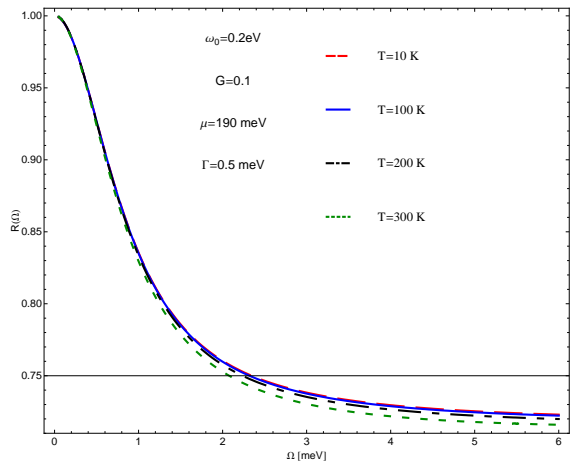}}
\caption{The ratio of dressed (by phonons) to bare conductivity $R(\Omega)$ defined in the text
vs $\Omega$ in the low-energy (to $\SI{6}{meV}$) at four temperatures:
$T = \SI{10}{K}$ -- the long dashed (red) line,  $T = \SI{100}{K}$ -- the solid (blue) line,
$T = \SI{200}{K}$ -- the dash-dotted (black) line and $T = \SI{300}{K}$ -- the short dashed (green) line.
} \label{fig:3}
\end{figure}
Numerical results for a value of $R(\Omega,T)$ are shown in Fig.~\ref{fig:3} for four temperatures
$T = \SI{10}{K}, \SI{100}{K}, \SI{200}{K}, \SI{300}{K}$. All curves start at value 1 and are close to
a saturation value of $\sim 0.72$ by $\Omega = \SI{6}{meV}$
with a small decrease in saturation value at higher temperature taken to mean that the effective phonon renormalization
$\lambda_{eff}$ has increased slightly with temperature. This is a well known effect in simple metals \cite{Grimvall.book}.
Of course, in an actual experiment one does not have an access to the bare band conductivity and the ratio $R(\Omega,T)$
cannot be formed. However, one can take ratios at different temperatures to verify that the electron-phonon renormalization is
nearly a constant in temperature as found in this work. What is more important is that Fig.~\ref{fig:3} shows clearly that
the intraband Drude type renormalizations of Eq.~(\ref{Drude-phonon}) are significant for the parameters used here and should
be observable.

While we have digressed somewhat from our main theme, to describe more general effects of phonons on the conductivity,
our important result is that, in principle, the electron-phonon interaction will reduce the value of the universal
background below its bare band value. Admittedly for the parameters characteristic of graphene this effect is not expected to be large.
We note that in Ref.~\cite{Katsnelson2008EPL} the conclusion was made that in general the effects of Fermi liquid corrections on
the universal background are small.

\section{AC conductivity in an external magnetic field without phonons}

As we already mentioned in the Introduction, in the papers \cite{Gusynin2006PRB,Gusynin2007PRL,Gusynin2007JPCM,Gusynin2007PRB}
we have derived and analyzed various expressions for the diagonal and off-diagonal magneto-optical conductivity
of graphene. The derivation is also based on the analytical calculation of the bubble diagram (\ref{Pi-via-RA}), but instead of
the GF (\ref{GF}) one has to substitute the GF describing Dirac fermions in an external magnetic field.

Particularly useful are the Eqs.~(11) and (12) of Ref.~\cite{Gusynin2007JPCM}.
The first of these equations is for the complex diagonal conductivity
\begin{eqnarray}
\label{sigma_xx-complex-corrected1}
\fl \sigma_{xx}(\Omega)=\frac{e^2v_F^2|eB|(\Omega+2i\Gamma)}{\pi
c\,i}  \\
\fl \times \sum_{n=0}^{\infty}
\left\{\left(1-\frac{\Delta^2}{M_nM_{n+1}}\right)\frac{[n_{F}(M_n) -
n_F(M_{n+1})] + [n_F(-M_{n+1}) -
n_F(-M_{n})]}{(M_{n+1}-M_n)^2-(\Omega+2i\Gamma)^2}\frac{1}{M_{n+1}-M_n}
\right. \nonumber \\
\fl
+\left.\left(1+\frac{\Delta^2}{M_nM_{n+1}}\right)\frac{[n_{F}(-M_n)
- n_F(M_{n+1})] + [n_F(-M_{n+1}) -
n_F(M_{n})]}{(M_{n+1}+M_n)^2-(\Omega+2i\Gamma)^2}\frac{1}{M_{n+1}+M_n}\right\},
\nonumber
\end{eqnarray}
where the energies of the relativistic
Landau levels are
\begin{equation}
\label{M_n} E_n=\pm M_n, \qquad
M_{n}=\sqrt{\Delta^{2}+2nv_F^2|eB|/c}.
\end{equation}
In Eq.~(\ref{M_n}) the Landau energy scale $L(B)$ associated with the magnetic field expressed
in the units of temperature and energy reads [restoring $\hbar$]
\begin{eqnarray}
\fl && L^2(B) = \frac{2e B v_F^2}{c} \to \frac{2eB \hbar v_F^2}{c k_B^2}  =
1.7696 \times 10^{-7} \mbox{K}^2 v_F^2 (\mbox{m/s}) B [\mbox{T}],  \\
\fl && L^2(B) \approx \SI{1314}{meV^2} B[\mbox{T}], \quad L(B) \approx \SI{36.3}{meV} \sqrt{B[\mbox{T}]}, \quad
E_n  \approx \pm   \sqrt{n}  L(B), \nonumber
\end{eqnarray}
where the field $B$ is given in Tesla and $v_F$ is the Fermi velocity in graphene given in m/s
and in the second line it is chosen to be $v_F \approx \SI{10^6}{m/s}$ [in what follows we set $\Delta =0$].
It is assumed in Eq.~(\ref{sigma_xx-complex-corrected1}) that all Landau levels have the same width,
for the case of the Landau levels with different widths see Eq.~(9) of \cite{Gusynin2007JPCM}.

To study the  limit of a weak magnetic field, we apply the Poisson summation formula
and express the conductivity (\ref{sigma_xx-complex-corrected1}) as follows
\begin{eqnarray}
\label{sigma-Poisson}
\fl {\rm Re}\sigma_{xx}(\Omega)=\frac{e^{2}}{4\pi\hbar}\left[\frac{1}{2}L^{2}f(0)+\int\limits_{0}^{\infty}
dx f(x)+\sum\limits_{k=1}^{\infty}\int\limits_{0}^{\infty}dx f(x) \cos\left(\frac{2\pi k x}{L^{2}}\right)\right],
\end{eqnarray}
where $f(x)=f^{intra}(x)+f^{inter}(x)$. The intraband and interband contributions are given by
\ba
\fl f^{intra}(x)&=&\frac{n_{F}(\sqrt{x})-n_{F}(\sqrt{x+L^{2}})+n_{F}(-\sqrt{x+L^{2}})
-n_{F}(-\sqrt{x})}{\sqrt{x+L^{2}}-\sqrt{x}}
\nonumber\\
\fl &\times&\left[\frac{2\Gamma}{\left(\sqrt{x+L^{2}}-\sqrt{x}-\Omega\right)^{2}+(2\Gamma)^{2}}
+\frac{2\Gamma}{\left(\sqrt{x+L^{2}}-\sqrt{x}+\Omega\right)^{2}+(2\Gamma)^{2}}\right],
\ea
and
\ba
\label{f-inter}
\fl f^{inter}(x)&=&\frac{n_{F}(-\sqrt{x})-n_{F}(\sqrt{x+L^{2}})+n_{F}(-\sqrt{x+L^{2}})
-n_{F}(\sqrt{x})}{\sqrt{x+L^{2}}+\sqrt{x}}
\nonumber\\
\fl &\times&\left[\frac{2\Gamma}{\left(\sqrt{x+L^{2}}+\sqrt{x}-\Omega\right)^{2}+(2\Gamma)^{2}}
+\frac{2\Gamma}{\left(\sqrt{x+L^{2}}+\sqrt{x}+\Omega\right)^{2}+(2\Gamma)^{2}}\right].
\ea
Although the representation (\ref{sigma-Poisson}) - (\ref{f-inter}) looks rather complicated, its limiting
case provide a lot of insight on the behaviour of the conductivity.
For example, evaluation of Eq.~(\ref{sigma-Poisson}) in the high frequency limit, $\Omega \gg L(B),|\mu|,\Gamma$, gives
\begin{eqnarray}
\label{sigma_ac-osc}
\fl && \mbox{Re}
\sigma_{xx}(\Omega) \simeq\\
\fl && \frac{e^{2}}{2h}\left\{\pi+\frac{4\Gamma}{\Omega}
\sum\limits_{k=1}^{\infty}\int\limits_{0}^{\infty}du\left[\frac{1}{(u+1)^{2}+(2\Gamma/\Omega)^{2}}+
\frac{1}{(u-1)^{2}+(2\Gamma/\Omega)^{2}}\right]
\cos\left(\pi k\frac{u^{2}\Omega^{2}}{2L^{2}(B)}\right)\right\}. \nonumber
\end{eqnarray}
The first term of Eq.~(\ref{sigma_ac-osc}) corresponds to the universal ac background, while the second term
describes  an oscillatory behavior about that value.
\begin{figure}[h]
\centering{
\includegraphics[width=0.48\textwidth]{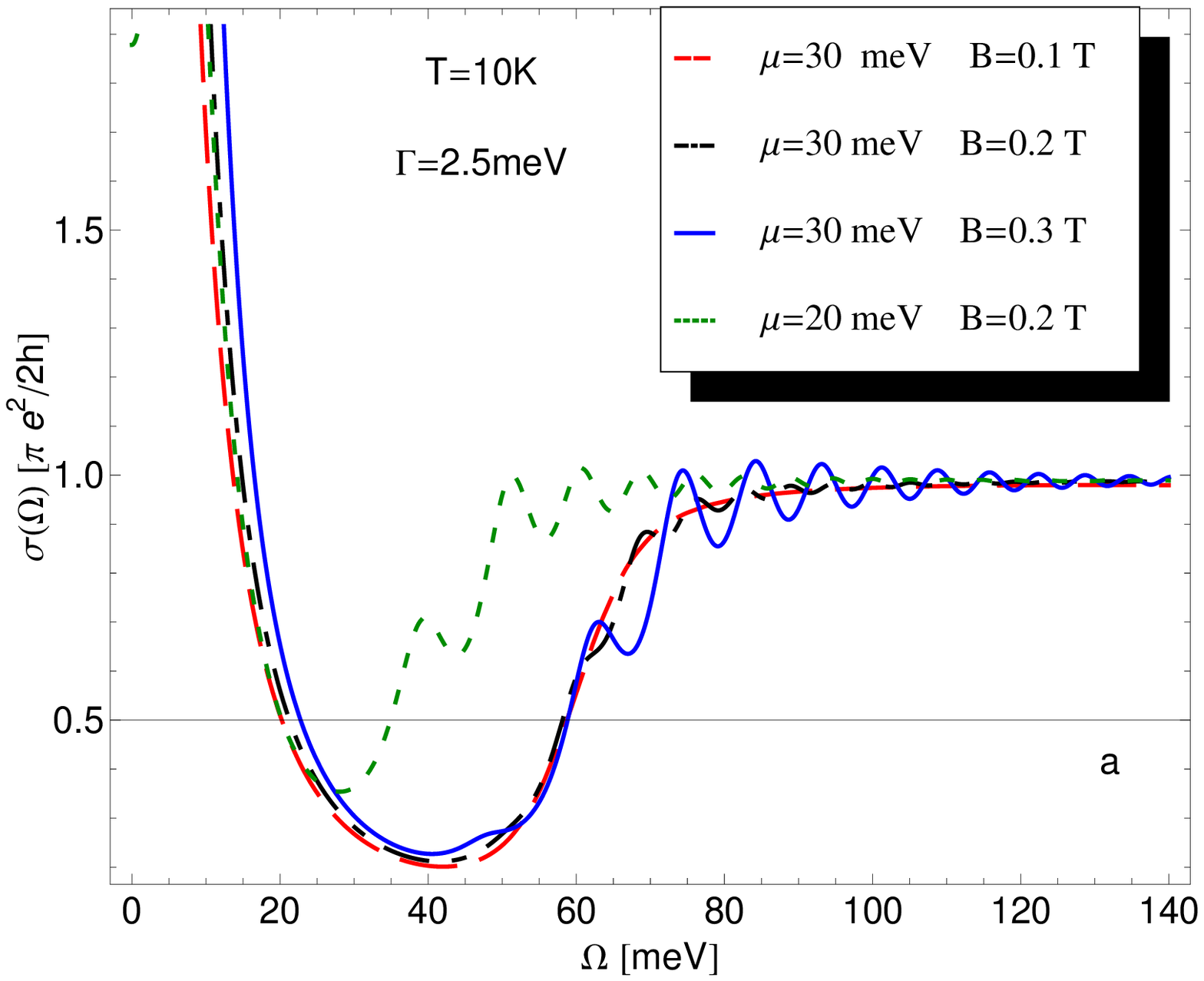}
\includegraphics[width=0.48\textwidth]{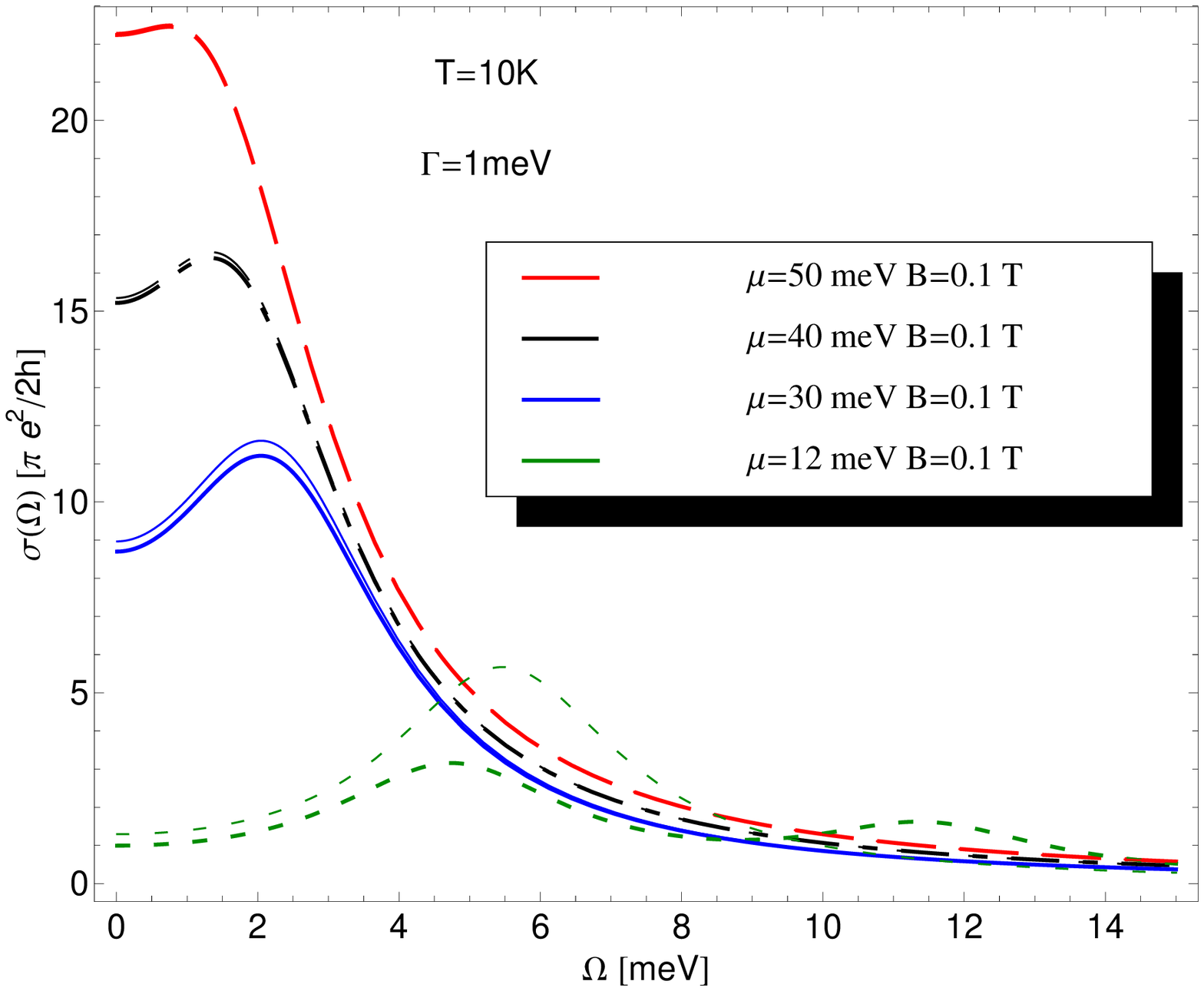}
}
\caption{(a) The conductivity $\mbox{Re} \sigma_{xx}(\Omega)$ in units $\pi e^2/2 h$ (universal background)
for temperature $T=\SI{10}{K}$, scattering rate $\Gamma = \SI{2.5}{meV}$.
Long dashed (red) line, the chemical potential
$\mu = \SI{30}{meV}$ and the magnetic field $B=\SI{0.1}{T}$,
dash-dotted (black) line $\mu =  \SI{30}{meV}$ and $B=\SI{0.2}{T}$, solid (blue) line
$\mu = \SI{30}{meV}$ and $B=\SI{0.3}{T}$, short dashed (green) line $\mu =\SI{20}{meV}$ and $B=\SI{0.2}{T}$.
(b)
The conductivity $\mbox{Re} \sigma_{xx}(\Omega)$ in units $\pi e^2/2 h$ (universal background)
for  magnetic field $B=\SI{0.1}{T}$, temperature $T=\SI{10}{K}$, and scattering rate $\Gamma = \SI{1}{meV}$.
Long dashed (red) line, the chemical potential
$\mu = \SI{50}{meV}$,
dash-dotted (black) $\mu =  \SI{40}{meV}$, solid (blue)
$\mu = \SI{30}{meV}$, short dashed (green) $\mu =\SI{10}{meV}$. All thick lines are computed using the full expression (\ref{sigma_xx-complex-corrected1}) and the thin lines using the first line of
Eq.~(\ref{sigma-low-B}).
}
\label{fig:4}
\end{figure}
In Fig.~\ref{fig:4}~a we illustrate this analytical result and present the
ac conductivity  $\mbox{Re} \sigma_{xx}(\Omega)$ in units $\pi e^2/2 h$
as a function of $\Omega$ to $\SI{150}{meV}$ computed using the formula
(\ref{sigma_xx-complex-corrected1}). It turns out that
even for the smallest value of the field $B=\SI{0.1}{T}$  considered it is
sufficient to include 100 terms in the sum. We observe indeed that for large $\Omega$ the conductivity
oscillates around the universal value.

In the clean limit, $\Gamma\to0$, and for fields $L (B) \gg |\mu|$ the main contribution to the integrals in
Eq.~(\ref{sigma-Poisson}) comes from the Lorentzians,  because they reduce to delta functions, and
thus we obtain
\ba
\label{sigma-high-B}
\fl \sigma_{xx} &&(\Omega)  =\frac{e^{2}}{4\hbar}\left\{ \left[n_{F}(-L)
-n_{F}(L)\right]\left[\frac{2\Gamma L}{\pi[(L-\Omega)^{2}+(2\Gamma)^{2}]}+
\frac{2\Gamma L}{\pi[(L+\Omega)^{2}+(2\Gamma)^{2}]}\right]
\right.\nonumber\\
\fl &&+\left.\left[n_{F}\left(\frac{L^{2}-\Omega^{2}}{2\Omega}\right)-
n_{F}\left(-\frac{L^{2}-\Omega^{2}}{2\Omega}\right)-n_{F}\left(\frac{L^{2}+\Omega^{2}}{2\Omega}\right)
+n_{F}\left(-\frac{L^{2}+\Omega^{2}}{2\Omega}\right)\right] \right. \nonumber \\
\fl && \times \left[\frac{|\Omega^{4}-L^{4}|}{2\Omega^{4}}
\right.\\
\fl &&+\left.\left.\frac{2}{\pi\Omega}\sum\limits_{k=1}^{\infty}
\int\limits_{0}^{\infty}dx\left[\frac{2\Gamma}{\left(\sqrt{x+L^{2}}-\sqrt{x}-\Omega\right)^{2}+(2\Gamma)^{2}}+
\frac{2\Gamma}{\left(\sqrt{x+L^{2}}+\sqrt{x}-\Omega\right)^{2}+(2\Gamma)^{2}}
\right]\right.\right.\nonumber\\
\fl &&\times\left.\left.\cos\left(\frac{2\pi k x}{L^{2}}\right)\right]\right\}, \nonumber
\ea
where we kept finite $\Gamma$ in the places crucial  for the regularization of the conductivity.
This high field regime $L(B) \gg |\mu|$ is, in principle, well described by the original representation
(\ref{sigma_xx-complex-corrected1}). For example, the first term of (\ref{sigma-high-B})
corresponds to the absorption line at $\Omega = E_1 - E_0 = L(B)$.
In the weak fields, $L (B) \lesssim |\mu|$, the intraband part has an additional sharp (at small temperatures)
maximum which gives the following contribution
\ba
&&\int\limits_{0}^{\infty}dx f(x)\simeq \int\limits_{0}^{\infty}dx
\left(-\frac{\partial n_{F}(\sqrt{x})}
{\partial\sqrt{x}}+\frac{\partial n_{F}(-\sqrt{x})}{\partial\sqrt{x}}\right)\nonumber\\
&&\times\left[\frac{2\Gamma}{\left(\frac{L^{2}}{2\sqrt{x}}-\Omega\right)^{2}+(2\Gamma)^{2}}
+\frac{2\Gamma}{\left(\frac{L^{2}}{2\sqrt{x}}+\Omega\right)^{2}+(2\Gamma)^{2}}\right].
\ea
Since the maximum of the integral is at the point $x=\mu^{2}$, one can replace $L^{2}/2\sqrt{x}$ by
the cyclotron frequency  $\omega_c = |eB|v_F^2/(c|\mu|)$\footnote{Accordingly, we have $\hbar \omega_c[\mbox{meV}] =
\SI{657}{meV^2}B[\mbox{T}] /\mu \mbox{[meV]}$}, then the remaining integrals are
evaluated exactly and we obtain
\begin{eqnarray}
\label{sigma-low-B}
\fl {\rm Re}\sigma_{xx}(\Omega)&=&\frac{2 e^2}{h}T\ln\left(2\cosh\frac{\mu}{2T}\right)
\left[\frac{2\Gamma}{(\omega_{c}-\Omega)^{2}+4\Gamma^{2}}+\frac{2\Gamma}{(\omega_{c}+
\Omega)^{2}+4\Gamma^{2}}\right] \nonumber\\
\fl &+& \frac{e^{2} L}{2h} \left[n_{F}(-L)
-n_{F}(L)\right]\left[\frac{2\Gamma}{(L-\Omega)^{2}+4\Gamma^{2}}+
\frac{2\Gamma}{(L+\Omega)^{2}+4\Gamma^{2}}\right]\\
\fl &+&   \frac{\pi e^2}{4h}\left[\tanh\frac{\Omega+2\mu}{4T}
+\tanh\frac{\Omega-2\mu}{4T}\right], \qquad E_1 = L(B) \lesssim |\mu|,\nonumber
\end{eqnarray}
where similarly to
Eqs.~(\ref{B=0.cond-intra-inter-T=0}) and (\ref{B=0.cond-intra-inter-T}),
the impurity scattering rate $\Gamma$ was set to zero in the interband term.
We note that setting $B=0$ in Eq.~(\ref{sigma-low-B}), we arrive at the
Eqs.~(13) and (14) from Ref.~\cite{Gusynin2007JPCM} mentioned in Sec.~\ref{sec:gap}.
Eq.~(\ref{sigma-low-B}) illustrates that the universal ac background is robust to the perturbations provided
by a weak magnetic field. For nonzero, but small $B$ such that $E_1 = L(B) \lesssim |\mu|$
(or equivalently, for $2 \hbar \omega_c \lesssim |\mu|$) the first line
of (\ref{sigma-low-B}) with the same coefficient $\ln\left[2\cosh(\mu/2T)\right]$ as in the $B=0$
expression (\ref{B=0.cond-intra-inter-T}), is much larger than the second line of (\ref{sigma-low-B}). In this case,
the spectral weight $W_{intra}$ given by Eq.~(\ref{spectral-weights})
of the Drude peak at $\Omega=0$ from Eq.~(\ref{B=0.cond-intra-inter-T}) is  transferred to the
cyclotron peak at $\Omega = \omega_c$ which has exactly the same field independent spectral weight.
Further, as the field increases, the dependent on this field spectral weight of the second line also grows and a second peak at
$\Omega = E_1$ develops. As the field grows and becomes $E_1 = L(B) \gtrsim |\mu|$, one has to use the appropriate
in this limit expression (\ref{sigma-high-B}) which has only one low-energy peak at $\Omega = E_1$.

In Fig.~\ref{fig:4}~b we show the evolution of the Drude peak to
the cyclotron peak at $\Omega = \omega_c$ observed when $E_1 \lesssim |\mu|$.
One can see that for $\mu = \SI{50}{meV}$ the thick long dashed (red) is undistinguishable
from the thin  long dashed (red) computed using the approximate expression based on the first line
of Eq.~(\ref{sigma-low-B}). For $\mu =  \SI{40}{meV}$ thick and thin dash-dotted (black)
lines are already distinguishable, but show that the agreement between the full expression (\ref{sigma_xx-complex-corrected1})
and the first line of Eq.~(\ref{sigma_xx-complex-corrected1}) is very good. The agrement remains good for lower value of
$\mu = \SI{30}{meV}$ [for solid (blue) lines]. However, when $\mu$ is getting close to $E_1 =\SI{11.5}{meV}$
as shown in the short dashed (green) curve for  $\mu =\SI{12}{meV}$, the approximation based on (\ref{sigma-low-B})
loses its validity. The position of the first peak on the thick short dashed line is shifted from $\Omega = \omega_c$
to lower energy and second peak at $\Omega = E_1$ develops.

\section{Discussion}

We have studied how the universal ac conductivity background in graphene is affected
by a finite chemical potential, finite temperature, elastic impurity scattering, application of a weak magnetic field,
as well as the opening of a finite gap which changes the quasiparticles into massive relativistic-like fermions.
While large changes in the ac conductivity are found to arise at low energies, when the photon energy is increased to a value
much larger than any of the energy scales involved, the corrections to the universal conductivity
coming from the above mentioned sources are small and the
background value remains close to $\pi e^2/(2h)$. On the other hand, the changes found when the electron-phonon interaction
is accounted for can be more significant for large values of coupling to high energy phonons.
These renormalizations increase the band width on the magnitude of the phonon energy scale and thus transfer optical weight
to states beyond the bare band cutoff and consequently reduce the background at intermediate energies. Of course, if the
actual electron-phonon coupling in graphene is very small these effects will also be small. Nevertheless, it is important
that the background value can be affected by various agencies and need not be exactly $\pi e^2/(2h)$.
The electron-phonon interaction also renormalizes the value of the
chemical potential measurable as a threshold in the absorption.

We find that this interaction, while providing phonon assisted sidebands called Holstein processes in metal physics,
does not introduce wiggles in the background value as confirmed in \cite{Peres2009EPL}.
We demonstrated that in a weak external magnetic field  the universal
value of the ac conductivity does not change while the conductivity itself oscillates around this value.
Finally, we studied the evolution of the Drude peak in graphene with a magnetic field.

\ack

We thank E.J.~Nicol for important discussions on electron-phonon effects.
S.G.S. thanks L.~Benfatto for stimulating discussion and L.A.~Falkovsky for comments
on an early version of the manuscript.
V.P.G. and S.G.S. were supported by  the special program
of the NAS of  Ukraine and by the Ukrainian State Foundation for Fundamental Research
under Grant No. F28.2/083. The work of V.P.G was also partially supported by the Grant
``Nanostructure systems, nanomaterials, nanotechnologies'' No. 10/07-N of the NAS of  Ukraine.
This work has been supported by NSERC  and by the
Canadian Institute for Advanced Research (CIFAR) (J.P.C.).


\section*{References}

\begin{thebibliography}{99}

\bibitem{Geim2007NatMat} Geim A K and Novoselov K S, 2007 {\it Nat. Mat.} {\bf 6} 183

\bibitem{Neto2009RMP} Castro Neto A H, Guinea F,  Peres N M R, Novoselov K S and Geim A K,
2009 {\it Rev. Mod. Phys.} {\bf 81} 109

\bibitem{Wang2008Science} Wang F, Zhang Y, Tian C, \etal
2008 {\it Science} {\bf 320} 206; Mak K F \etal 2008 {\it Phys. Rev. Lett.} {\bf 101}  196405

\bibitem{Nair2008Science} Nair R R, Blake P, Grigorenko A N, \etal
2008 {\it Science} {\bf 320} 1308

\bibitem{Li2008NatPhys} Li Z Q, Henriksen E A, \etal
2008 {\it Nature Phys.} {\bf 4} 532

\bibitem{Kuzmenko2009PRB} Kuzmenko A B,  van Heumen E, van der Marel D, \etal
2009 {\it Phys. Rev. B} {\bf 79} 115441

\bibitem{Dawlaty2008APL} Dawlaty J M, Shivaraman S,  Strait J, \etal
2008 {\it Appl.~Phys.~Lett.}  {\bf 93} 131905

\bibitem{Kuzmenko2008PRL} Kuzmenko A B, van Heumen E, Carbone F and van der Marel D,
2008 {\it Phys.~Rev.~Lett.}  {\bf 100} 117401

\bibitem{Ando2002JPSJ} Ando T, Zheng Y and Suzuura H
2002 {\it J. Phys. Soc. Jpn.} {\bf 71} 1318

\bibitem{Gusynin2006PRL} Gusynin V P, Sharapov S G and Carbotte J
P 2006 {\it Phys.~Rev.~Lett.} {\bf 96} 256802

\bibitem{Falkovsky2007EPJB} Falkovsky L A and Varlamov A A 2007 {\it Eur. Phys. J. B}
{\bf 56} 281

\bibitem{Falkovsky2007PRB} Falkovsky L A and Pershoguba S S 2007 {\it Phys. Rev. B}
{\bf 76} 153410

\bibitem{Stauber2008aPRB} Stauber T,  Peres N M R and Geim A K 2008
{\it Phys. Rev. B} {\bf 78} 085432

\bibitem{Sadowski2006PRL} Sadowski M L, Martinez G, Potemski M, Berger C and
de~Heer W A 2006 {\it Phys. Rev. Lett.} {\bf 97} 266405

\bibitem{Plochocka2008PRL} Plochocka P, Faugeras C, Orlita M, \etal
2008 {\it Phys. Rev. Lett.} {\bf 100} 087401

\bibitem{Jiang2007PRL} Jiang Z, Henriksen E A, Tung L C,  \etal
2007 {\it Phys. Rev. Lett.} {\bf 98} 197403

\bibitem{Deacon2007PRB} Deacon R S, Chuang K C, Nicholas R J, \etal
2007 {\it Phys. Rev. B} {\bf 76} 081406

\bibitem{Henriksen2008PRL} Henriksen E A, Jiang Z, Tung  L C,  \etal
2008 {\it Phys. Rev. Lett.} {\bf 100} 087403

\bibitem{Li2006PRB} Li Z Q, Padilla W J, Tsai S-W, Dordevic S V,
Burch  K S, Wang Y J, and Basov D N 2006
{\it Phys. Rev. B} {\bf 74} 195404

\bibitem{Peres2006PRB} Peres N M R, Guinea F and Castro Neto A H
2006 {\it Phys. Rev. B} {\bf 73} 125411

\bibitem{Gusynin2006PRB} Gusynin V P and Sharapov S G 2006
{\it Phys. Rev. B} {\bf 73} 245411

\bibitem{Gusynin2007PRL} Gusynin V P, Sharapov S G and Carbotte J
P 2007 {\it Phys.~Rev.~Lett.} {\bf 98} 157402

\bibitem{Gusynin2007JPCM} Gusynin V P, Sharapov S G and Carbotte J P
2007 {\it J. Phys. Cond. Matt.} {\bf 19} 026222

\bibitem{Bychkov2008PRB} Bychkov Yu A and Martinez G 2008
{\it Phys. Rev. B} {\bf 77} 125417

\bibitem{Fischer2009} Fischer A M, Dzyubenko A B and R\"{o}mer R A
2009 Preprint arXiv:0902.4176


\bibitem{Peres2008EPL} Peres N M R, Stauber T and Castro Neto A H 2008
{\it Europhys. Lett.} {\bf 84} 38002

\bibitem{Stauber2008PRB} Stauber T,  Peres N M R and Castro Neto A H 2008
{\it Phys. Rev. B} {\bf 78} 085418

\bibitem{Stauber2008JPCM} Stauber T and  Peres N M R 2008
{\it J. Phys. Cond. Matt.} {\bf 20} 055002

\bibitem{Mishchenko2007PRL} Mishchenko E G 2007 {\it Phys. Rev. Lett.} {\bf 98} 216801;
Mishchenko E G 2008 {\it Europhys. Lett.} {\bf 83} 17005

\bibitem{Herbut2008PRL} Herbut I F, Juricic V and Vafek O
2008 {\it Phys. Rev. Lett.} {\bf 100} 046403

\bibitem{Fritz2008PRB} Fritz L, Schmalian J, Mueller M and Sachdev S
2008 {\it Phys. Rev. B} {\bf 78} 085416

\bibitem{Sheehy2009} Sheehy D E and Schmalian J 2009 Preprint arXiv:0906.5164

\bibitem{Basov2005RMP} Basov D N and Timusk T 2005 {\it Rev. Mod. Phys.} {\bf 77} 721


\bibitem{Carbotte:review} Carbotte J P and Schachinger E 2006 {\it J. Low Temp. Phys.} {\bf 144} 61

\bibitem{Benfatto:review} Benfatto L and Sharapov S  2006 {\it Fiz. Nizk. Temp.} {\bf 32} 700
[2006 {\it Low Temp. Phys.} {\bf 32} 533]

\bibitem{Gusynin2007PRB} Gusynin V P, Sharapov S G and Carbotte J P 2007
{\it Phys.~Rev.~B} {\bf 75} 165407


\bibitem{Wallace1947PRev} Wallace R P 1947
{\it Phys.~Rev.} {\bf 77} 622

\bibitem{Sabio2008PRB} Sabio J, Nilsson J, and Castro Neto A H
2008 {\it Phys.~Rev.~B} {\bf 78} 075410

\bibitem{Semenoff1984PRL} Semenoff G W 1984
{\it Phys.~Rev.~Lett.}  {\bf 53}  2449

\bibitem{Gusynin2007IJMPB}
Gusynin V P, Sharapov S G, and Carbotte J P 2007
{\it Int. J. Mod. Phys. B} {\bf 21} 4611

\bibitem{Cappelluti2009PRB} Cappelluti E and Benfatto L 2009 {\it  Phys. Rev. B} {\bf 79} 035419


\bibitem{Falkovsky2008UFN} Falkovsky L A 2008 {\it Physics-Uspekhi} {\bf 51} 887

\bibitem{Peres2008IJMPB} Peres N M R and Stauber T 2008 Int. J. Mod. Phys. B {\bf 22} 2529

\bibitem{Benfatto2008PRB} Benfatto L and  Cappelluti E 2008 {\it Phys. Rev. B} {\bf 78} 115434


\bibitem{Park2007PRL} Park C-H, Giustino F, Cohen M L and Louie S G 2007 {\it Phys. Rev. Lett.} {\bf 99} 086804

\bibitem{Calandra2007PRB} Calandra M and Mauri F 2007 {\it Phys. Rev. B} {\bf 76} 205411

\bibitem{Tse2007PRL} Tse W-K and Das Sarma S
2007 {\it Phys. Rev. Lett.} {\bf 99} 236802

\bibitem{Bostwick2007Nature} Bostwick A, Ohta T, Seyller T, Horn K and Rotenberg E 2007
{\it Nature Phys.} {\bf 3} 36

\bibitem{Li2009PRL} Li G, Luican A and Andrei E Y 2009 {\it Phys. Rev. Lett.} {\bf 102} 176804

\bibitem{Carbotte2009} Carbotte J P, Nicol E J, and Sharapov S G 2009 arXiv:0908.2608

\bibitem{Peres2009EPL} Peres N M R, Stauber T and Castro Neto A H 2009
{\it Europhys. Lett.} {\bf 86} 49901; Stauber T, Peres N M R and A. H. Castro Neto 2009 {\it Phys. Rev. B} {\bf 79} 239901(E)


\bibitem{Grimvall.book} Grimvall G 1981 {\it The Electron-Phonon Interaction in Metals} (New York, North-Holland)

\bibitem{Katsnelson2008EPL} Katsnelson M I {\it Europhys. Lett.} 2008 {\bf 84} 37001



\end{thebibliography}
\end{document}